\documentclass[twocolumn,english,superscriptaddress,amssymb,noshowpacs]{revtex4}
\usepackage[T1]{fontenc}
\usepackage[latin9]{inputenc}
\usepackage{color}
\usepackage{amsmath}
\usepackage{amssymb}
\usepackage{graphicx}

\makeatletter

\@ifundefined{textcolor}{}
{%
 \definecolor{BLACK}{gray}{0}
 \definecolor{WHITE}{gray}{1}
 \definecolor{RED}{rgb}{1,0,0}
 \definecolor{GREEN}{rgb}{0,1,0}
 \definecolor{BLUE}{rgb}{0,0,1}
 \definecolor{CYAN}{cmyk}{1,0,0,0}
 \definecolor{MAGENTA}{cmyk}{0,1,0,0}
 \definecolor{YELLOW}{cmyk}{0,0,1,0}
}

\usepackage{bm}\usepackage{color}

\usepackage[english]{babel}

\makeatother

\usepackage{babel}
\usepackage{mathrsfs}

\begin{document}

\title{Quantum dynamics of an electromagnetic mode that cannot \textcolor{black}{contain} $N$ photons}

\author{L. Bretheau}
\author{P. Campagne-Ibarcq}
\author{E. Flurin}
\author{F. Mallet}
\author{B. Huard}
\thanks{Corresponding author. E-mail: benjamin.huard@ens.fr}
\affiliation{Laboratoire Pierre Aigrain, Ecole Normale Supérieure-PSL Research University, CNRS,
Université Pierre et Marie Curie-Sorbonne Universités, Université Paris Diderot-Sorbonne Paris Cité,
\\24 rue Lhomond, 75231 Paris Cedex 05, France}

%

\begin{abstract}
Electromagnetic modes are instrumental in 
building quantum machines. \textcolor{black}{In this experiment, we introduce a method to manipulate \textcolor{black}{these modes} by effectively controlling their phase space.} Preventing access to a single energy level, corresponding to a number of photons $N$, confined the dynamics of the field to levels $0$ to $N-1$. 
Under a resonant drive, the level occupation was found to oscillate in time, similarly to an $N$-level system. 
Performing a direct Wigner tomography of the field revealed its nonclassical features, including a Schr\"odinger cat-like state at half period in the evolution. This \textcolor{black}{fine} control of the field in its phase space may enable applications in quantum information and metrology.
\end{abstract}

\maketitle

The manipulation of a quantum system usually involves the control of its Hamiltonian in time. An alternative route consists in effectively tailoring its Hilbert space dynamically. This can be done by restricting the \textcolor{black}{system} evolution to a subset of possible states. 
When even a single energy level is disabled, the system evolution is deeply modified and is ruled by the so-called quantum Zeno dynamics (QZD)~\cite{Facchi2000,Facchi2002,Facchi2004,Raimond2010,Raimond2012}. \textcolor{black}{As the name suggests,} the level blockade can be realized by repeatedly checking whether the level is occupied\textcolor{black}{, owing to the inherent back action of quantum measurements}~\cite{Misra1977,Facchi2000,Facchi2002}. Alternatively, as in \textcolor{black}{the present} experiment, QZD can be achieved by blocking the level using a strong, active coupling to an ancillary quantum system~\cite{Facchi2004,Raimond2010,Raimond2012}, without any measurement~~\cite{Footnote1}.
\textcolor{black}{These} ideas have recently been demonstrated \textcolor{black}{for} atoms, using either Rb Bose-Einstein condensates~\cite{Schafer2014} or Rydberg atoms~\cite{Signoles2014}. \textcolor{black}{However,} the dynamics of these systems is intrinsically confined to a finite number of energy levels.
Here, \textcolor{black}{using a circuit quantum electrodynamics architecture, }we implement QZD of light.
With its large number of energy levels and ease of control, a single electromagnetic mode offers a wider and more controllable phase space than atoms and two-level systems.

\begin{figure}[h!b]
\includegraphics[width=1\columnwidth]{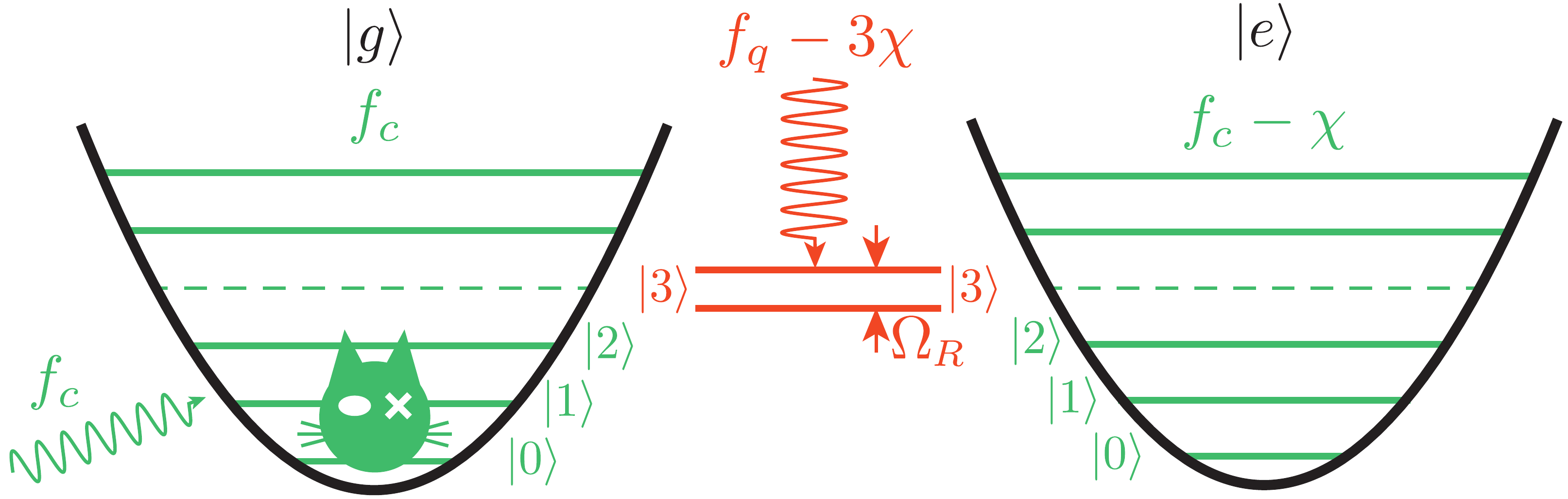}
\caption{\textbf{Principle of the experiment.}
Combined energy level diagram of the cavity coupled to the ancillary qubit. Each state is labelled as $|k\rangle\otimes|g/e\rangle$ for $k$ photons in the cavity and the qubit either in the ground ($g$) or excited ($e$) states.
The level $N=3$ is dynamically blocked by driving the qubit transition at $f_q-N\chi$ (red wave). \textcolor{black}{This drive} hybridizes the states $|N\rangle\otimes|g\rangle$ and $|N\rangle\otimes|e\rangle$ and shifts their energy. A coherent excitation at the cavity frequency $f_c$ (green wave) induces a periodic  evolution of the field confined to the first $N$ levels. At half period, a nonclassical state similar to a Schr\"odinger cat is produced. 
\label{fig1}}
\end{figure}

 The principle of our experiment is shown in Fig.~\ref{fig1}. One \textcolor{black}{cavity mode of} frequency $f_c$ is coupled to a qubit of frequency $f_q$. For large enough detuning, their evolution can be described by the dispersive Hamiltonian $hf_ca^\dagger a+hf_q|e\rangle\langle e|-h\chi a^\dagger a |e\rangle\langle e|$, where $h$ is Planck's constant, $a^\dagger$ is the ladder operator and $|e\rangle$ is the excited state of the qubit. The last term describes the frequency shift of the cavity (qubit) by $-\chi$, which occurs when the qubit (cavity) is excited by one extra quantum of energy. Owing to this shift, a tone at frequency $f_q-N\chi$ addresses only the transition between states $|N\rangle\otimes|g\rangle$ and $|N\rangle\otimes|e\rangle$ for level widths smaller than $\chi$~\cite{Schuster2007}; here $|g\rangle$ is the ground state of the qubit. These levels then hybridize and repel each other. Their splitting is given by the Rabi frequency $\Omega_R$ at which the qubit population would oscillate in \textcolor{black}{the case where} the cavity is in state $|N\rangle$ (Fig.~\ref{fig1}). Any transition to level $N$ is now forbidden when the cavity is driven at resonance. Schematically, level N has been moved out of the harmonic ladder (Fig.~\ref{fig1}). Then, starting from the ground state, the electromagnetic mode is confined to levels $0$ to $N-1$, whereas the qubit remains in its ground state. The field dynamics is dramatically changed\textcolor{black}{, resembling} that of an $N$-level system, and nonclassical states similar to Schr\"odinger cat states develop.

In the experiment, we use the fundamental mode of a three-dimensional (3D) microwave cavity made out of bulk aluminum, which resonates at $f_c=7.804~\mathrm{GHz}$. \textcolor{black}{This mode} is \textcolor{black}{off-resonantly} coupled to a superconducting qubit~\cite{Paik2011} with bare frequency $f_q=5.622~\mathrm{GHz}$ and dispersive frequency shift $\chi=4.63~\mathrm{MHz}$. The cavity exit rate $\gamma_c=(1.3~\mu\mathrm{s})^{-1}$ is dominated by the coupling rate to two transmission lines connected to the cavity, which are used for both driving and the readout of the system. The relaxation rate $\gamma_1=(11.5~\mu\mathrm{s})^{-1}$ and decoherence rate  $\gamma_2=(8.9~\mu\mathrm{s})^{-1}$ of the ancillary qubit are an order of magnitude smaller.

The experiment is performed by first turning on the blocking tone at $f_q-N\chi$. For the level blockade to be effective, we \textcolor{black}{choose} $\Omega_R=6.24~\mathrm{MHz}$ much larger than the level frequency width, which is about $\gamma_c$. Then the cavity is driven at frequency $f_d\approx f_c$ for a time $t$ varying up to a few $\mu\mathrm{s}$. The drive power is fixed throughout the experiment and would lead to \textcolor{black}{an amplitude displacement rate $\epsilon_d$ of about $3~\mu \mathrm{s}^{-1}$ in the cavity}, were there neither damping nor nonlinearities. At time $t$, both the blocking signal and the cavity drive are turned off, and the field state is measured. Two measurement schemes \textcolor{black}{are} used to characterize the cavity state. Both methods use as a probe the same qubit that is used to provide the level blockade.

\textcolor{black}{The first method}
consists in measuring the probability $P_k$ for the field to host $k$ photons. To do so, a selective $\pi$ pulse is applied to the qubit at frequency $f_q-k\chi$ so that it gets excited if $k$ photons are in the cavity. Measuring the probability to find the qubit in the excited state hence gives $P_k$~\cite{Kirchmair2013}. This direct measurement relies on the qubit being initially in the ground state, which is not always the case. To account for this, two spurious effects need to be considered. First, the qubit has a residual thermal population of 22~\%. Second, imperfections of the \textcolor{black}{blockade} induce a parasitic excitation of the qubit, which was measured in time and remains below 18~\%. Taking these effects into account leads to an accurate determination of the conditional 
probability $\tilde{P}_k$ of measuring $k$ photons\textcolor{black}{, given the qubit being initially in the ground state}~cite{Footnote2}.

\begin{figure}[h!t]
\includegraphics[width=1\columnwidth]{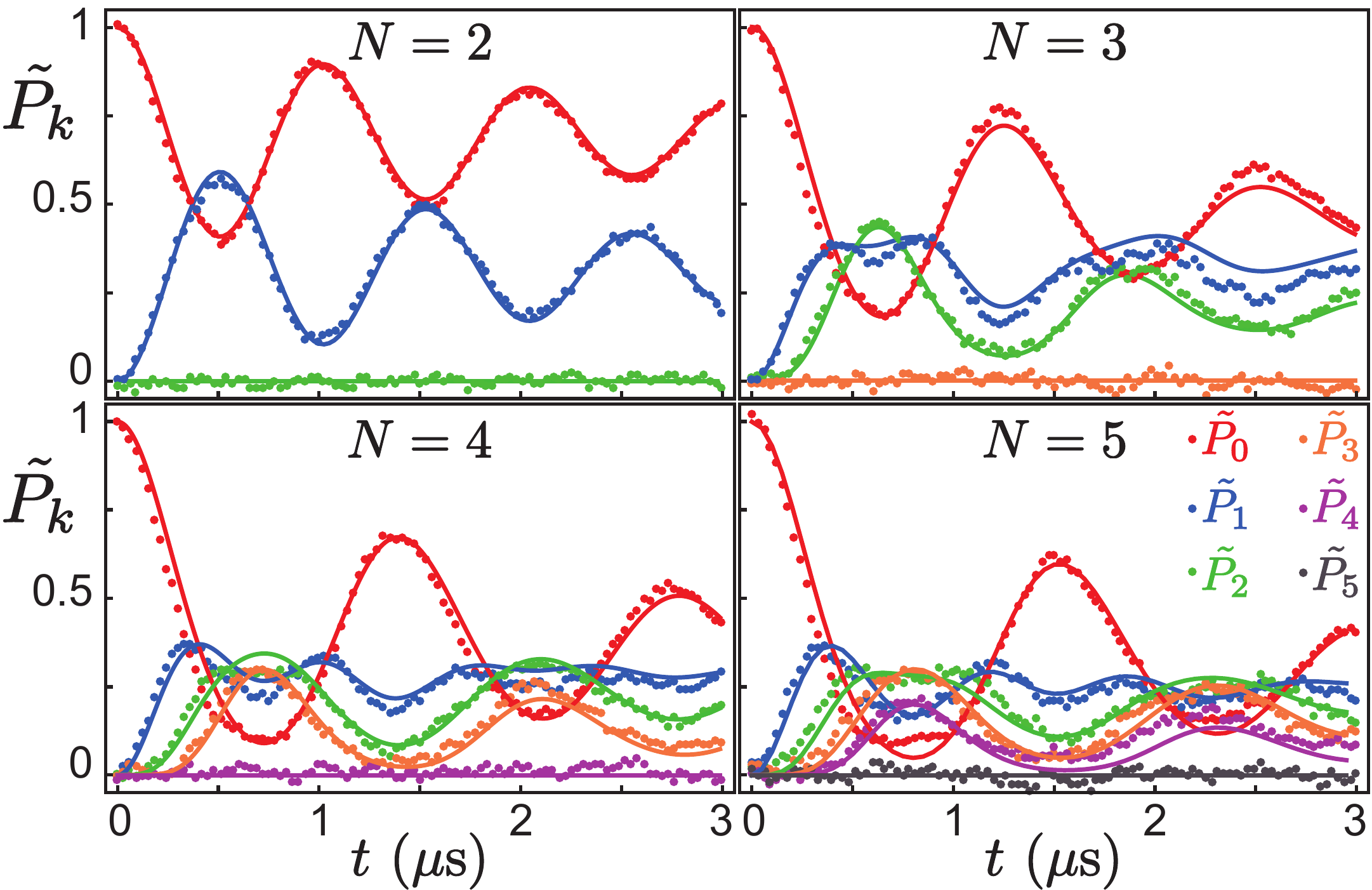}
\caption{\textbf{Evolution of the photon number probabilities.} Measured (dots) and theoretical (solid lines) photon number state probabilities $\tilde{P}_k$ as a function of time $t$. The blocked level $N$ ranges from 2 to 5 and is indicated on each panel. The standard deviation, estimated from the curve $\tilde{P}_{N}(t)$, is below $1.7 \%$. 
\label{figPk}}
\end{figure}

The resulting probabilities are shown in Fig.~\ref{figPk} as a function of time for several photon numbers $k$ and for $N$ from 2 to 5. Levels with more than $N$ photons are unoccupied, as expected from \textcolor{black}{Hilbert space blockade}.
Early in the evolution, the \textcolor{black}{occupation of the levels $k \ge 1$ rises in the} order of increasing energy, similarly to a coherent state of increasing amplitude (Fig.~S3c in~\cite{Footnote2}). \textcolor{black}{Later}, the level distribution \textcolor{black}{"bounces" off} a wall at $k=N$, so that the probabilities start to oscillate.
The period of the oscillations increases with $N$ as expected, because it takes more time to reach $N-1$ photons with a constant drive as $N$ increases.
The case $N=2$ is straightforward as it implements an effective qubit~\cite{Schafer2014}. The time traces of Fig.~\ref{figPk} correspond to Rabi oscillations of a two-level system. For larger $N$, the evolution is similar to that of a resonantly driven $N$-level system, as seen in Rydberg atoms~\cite{Signoles2014}. In particular, at half period, $\tilde{P}_0$ displays plateaus all the more pronounced as $N$ gets larger. Finally, the $N-1$ level occupation evolves in opposition to level 0, with a maximum at half period.

\begin{figure*}
\includegraphics[width=2.1\columnwidth]{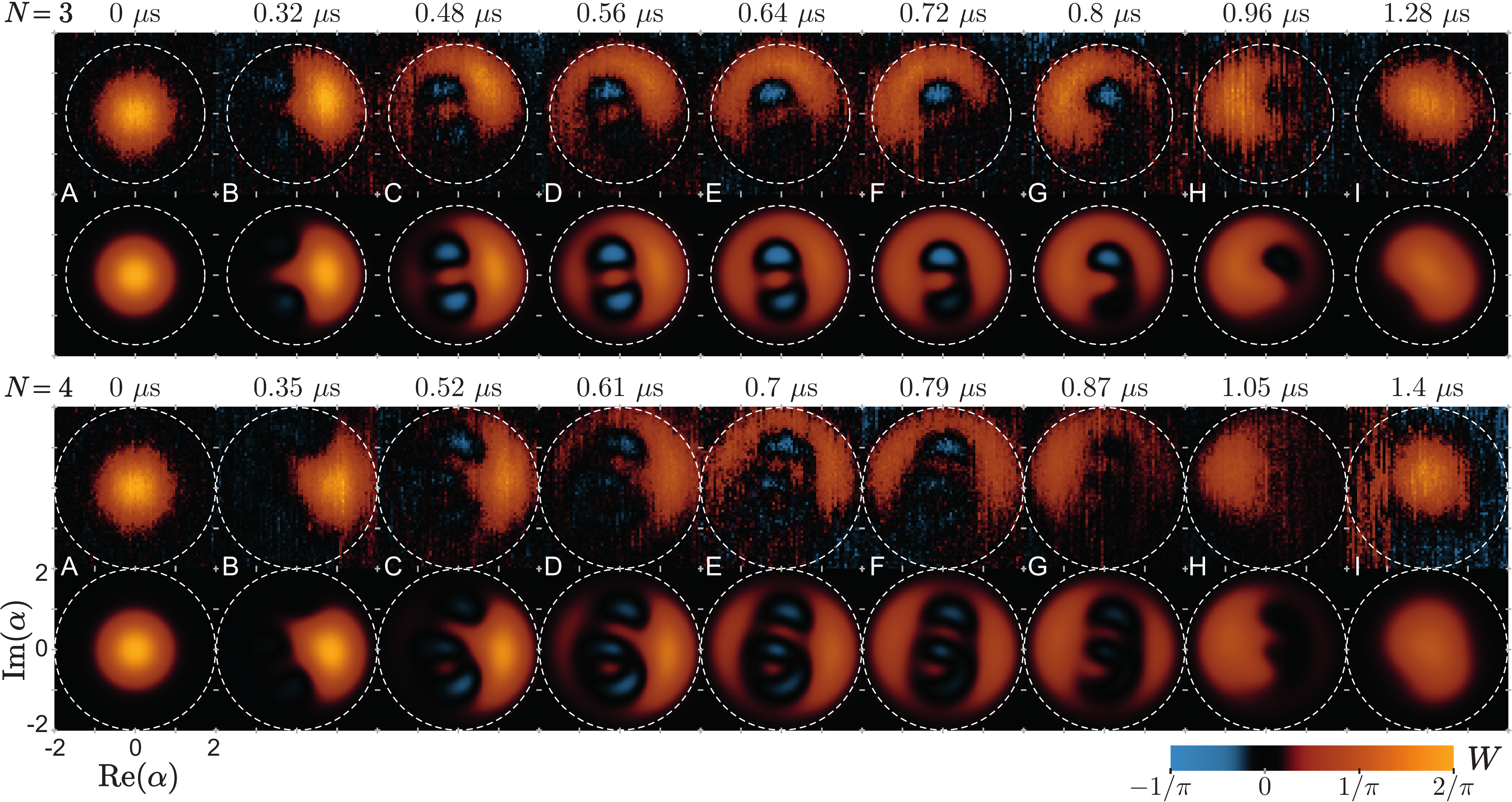}
\caption{\textbf{Time evolution of the Wigner function of a quantum mode of light under QZD.} Measured (top rows) and predicted (bottom rows) Wigner functions $W(\alpha)$ as a function of the displacement amplitude $\alpha$, for a blockade at $N=3$ (top panel) and $N=4$ (bottom panel). The time $t$ (nonlinear scale) for the frames shown in \textbf{A} to \textbf{I} is given above each panel. The field is confined in phase space by a barrier at amplitude $|\alpha|=\sqrt{N}$ (white dashed circle).
Negative values of the Wigner function, in blue, demonstrate the nonclassical nature of the field produced under QZD. The model used is the same as for Fig.~\ref{figPk}. Wigner functions are directly measured here and not reconstructed.
\label{fig3}}
\end{figure*}

The evolution of the field can be modeled with the Hamiltonian of the cavity
\begin{equation}
H= ih\epsilon_d(a_N^\dagger-a_N)/2\pi+h (f_c-f_d)a_N^\dagger a_N-h\lambda (a_N^\dagger)^2a_N^2,\label{H}
\end{equation}
written in the frame rotating at the drive frequency $f_d$.
These three terms describe the coherent drive, its detuning, and the nonlinearity of the cavity, respectively. The blockade at level $N$ suppresses transitions to and from $|N\rangle$. This is entirely modeled by an effective annihilation operator
\begin{equation}
a_N=a-\sqrt{N}|N-1\rangle\langle N|-\sqrt{N+1}|N\rangle\langle N+1|,
\end{equation}
for which these forbidden transitions are removed. Finally, the relaxation at rate $\gamma_c$ is modeled using a master equation of Lindblad form~\cite{Footnote2}.

Using this model, one can reproduce with good agreement the photon number probabilities $\tilde{P}_k$ (Fig.~\ref{figPk}), with the \textcolor{black}{displacement rate} $\epsilon_d$, the \textcolor{black}{drive} detuning $f_c-f_d$, and the cavity anharmonicity $\lambda$ as the fit parameters. 
\textcolor{black}{The displacement rate} $\epsilon_d$, which directly determines the period of oscillations in the \textcolor{black}{level} occupation (Fig.~\ref{figPk}), is found to be finely tuned between 2.83 and $3.05~\mu\mathrm{s}^{-1}$ for each blocked level $N$. The anharmonicity $\lambda=-70~\mathrm{kHz}$ is negative, as confirmed by independent measurements (Fig.~S2 in~\cite{Footnote2}). This negativity is in contradiction to all reported transmon qubit and 3D transmon models such as those in Refs.~\cite{Nigg2012,Bourassa2012,Solgun2014a}. At last, we attribute the origin of the detuning $f_c-f_d$ to an energy shift of the levels caused by the blocking tone. It was found to be $-0.1~\mathrm{MHz}$ for $N>2$ and $-0.4~\mathrm{MHz}$ for $N=2$. These values are consistent with the system being more strongly disturbed by the blocking field at $N=2$ than for larger $N$. 

These measurements demonstrate how the cavity is transformed into an $N$-level system by \textcolor{black}{dynamically inducing} a blockade at an arbitrary level $N$.
\textcolor{black}{A full characterization of the field, which goes beyond measuring only photon number probability, requires tomography. The Wigner function is a complete representation of the quantum state of the field} in continuous variables.
It can be expressed as $W(\alpha)=\langle D_{\alpha} \mathcal{P} D^\dagger _{\alpha}\rangle$, where  $D_\alpha=e^{\alpha a^\dagger - \alpha^* a}$ is the field displacement operator and $\mathcal{P}=e^{i\pi a^\dagger a}$ the photon parity operator.
A direct Wigner tomography of the field is \textcolor{black}{thus} performed by first displacing it by a complex amplitude $\alpha$ and then measuring the average photon parity. \textcolor{black}{This can be}  realized by mapping even and odd photon numbers on the qubit ground and excited states~\cite{Lutterbach1997,Bertet2002,Vlastakis2013}.
In practice, this is performed in three steps. 
The qubit is first prepared in state $(|g\rangle+|e\rangle)/\sqrt{2}$ using a pulse at the qubit frequency.
The $\pi/2$ pulse bandwidth is designed \textcolor{black}{to be} larger than $10\chi$ so as to succeed \textcolor{black}{regardless of} the cavity state.
Then, the qubit  \textcolor{black}{is left to} evolve \textcolor{black}{freely} during a time $\tau$, \textcolor{black}{where} it acquires a phase shift that depends on the cavity state owing to the evolution operator $e^{i2\pi\chi\tau a^\dagger a|e\rangle\langle e|}$. The waiting time is chosen \textcolor{black}{to be} $\tau=1/2\chi=108$~ns, so that the qubit phase increases by $k\pi$ when the cavity state is $|k\rangle$ (Fig.~S4 in~\cite{Footnote2}). 
Finally, another broadband $\pi/2$ pulse is applied in order \textcolor{black}{to flip} the qubit to the excited (ground) state in case of an even (odd) number of photons.
Measuring the qubit state then directly leads to the Wigner function in contrast to indirect reconstruction procedures such as maximum likelihood.

The measured Wigner functions for the field state are shown in Fig.~\ref{fig3} at various times during the first oscillation period for blocked levels $N=3$ and $N=4$. 
At time zero (Fig.~\ref{fig3}A), the cavity is in the vacuum state and the Wigner function is a positive Gaussian distribution centered  \textcolor{black}{at} $\alpha=0$, with a width reflecting zero point fluctuations.
As time increases, the resonant drive displaces the Wigner function along the real and positive axis.
Without the blockade, and neglecting losses and anharmonicity, the vacuum state observed at time $0$ would simply get displaced  \textcolor{black}{by} $\alpha=\epsilon_dt$.
In \textcolor{black}{the} presence of the blockade at level $N$, all levels with $k\geq N$ remain empty so that the field cannot be in a coherent state, and the distribution gets distorted.
Indeed in Fig.~\ref{fig3}B, the quasi-probability distribution seems to hit a wall in phase space at an amplitude $\sqrt{N}$. There, it undergoes a rapid counterclockwise evolution (Fig.~\ref{fig3}C to H) along the \textcolor{black}{corresponding} barrier (white dashed circle).
After a full oscillation period (Fig.~\ref{fig3}I), the cavity goes back to a state close to the vacuum  \textcolor{black}{state  of Fig.~\ref{fig3}A}. These snapshots of the field state are close to the ones predicted in Ref~\cite{Raimond2012} for light under QZD, and similar to the one observed in the levels of a Rydberg atom~\cite{Signoles2014}.

\begin{figure}
\includegraphics[width=1\columnwidth]{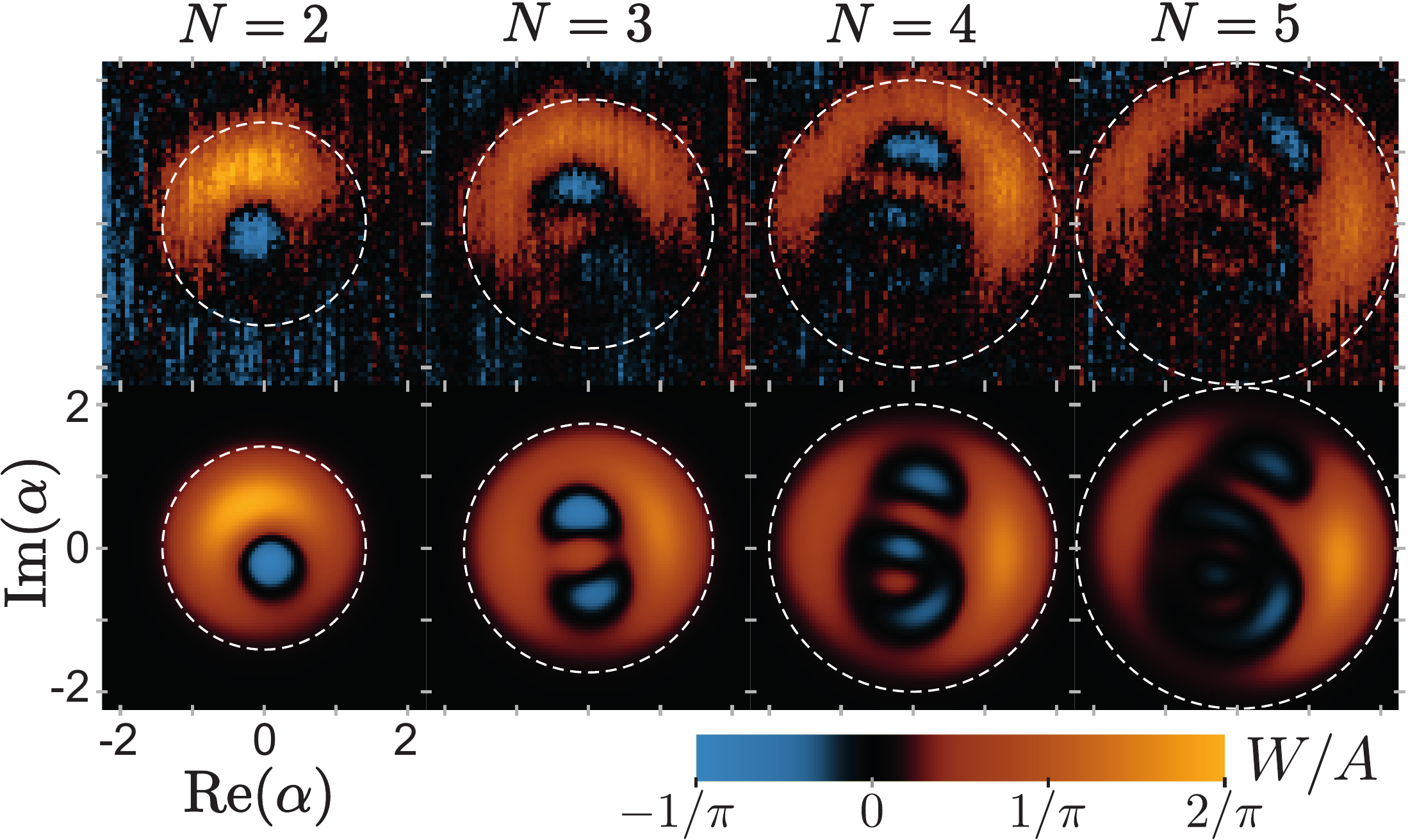}
\caption{\textbf{Wigner tomography at half period.} Measured (top row) and calculated (bottom row) Wigner functions of the cavity field for various blockade levels $N$ from $2$ to $5$, taken at half period ($t=0.51~\mu\mathrm{s},0.64~\mu\mathrm{s},0.7~\mu\mathrm{s}$ and $0.75~\mu\mathrm{s}$). The color scale is rescaled compared with Fig.~\ref{fig3} by $A=0.7$. Similarly to Schrödinger cat states, these states exhibit fringes with alternating positive and negative values.
\label{fig4}}
\end{figure}

Besides confirming the confined and periodic evolution of the field under QZD, this tomography reveals the formation of nonclassical field states, as indicated by the appearance of negative values in the Wigner function. These negative values develop while the field undergoes a transition along the  \textcolor{black}{barrier in phase space}. At half period, the state is close to an equal superposition of positive and negative field amplitudes at $\pm \sqrt {N-1}$. The quantum nature of this superposition manifests itself by the appearance of interference fringes in the Wigner function. These characteristic patterns can be compared for all blockade levels $2\leq N\leq 5$ in  Fig.~\ref{fig4}. The photon parity is that of the highest allowed level $N-1$, which is also the number of negative fringes. These properties are reminiscent \textcolor{black}{of} a quantum superposition between two coherent states of amplitude $\pm \sqrt {N-1}$, so-called Schr\"odinger cat state~\cite{Haroche2006}. The evolution of the Wigner function can be predicted using the model above. The calculated distributions are shown below each measurement frame in Figs.~\ref{fig3} and \ref{fig4}. The observed asymmetry between positive and negative imaginary amplitudes is attributed to drive detuning, losses and anharmonicity~\cite{Haroche2006,Kirchmair2013}. The model reproduces qualitatively the measured Wigner functions for all times and values of $N$. 

In conclusion, we have demonstrated Quantum Zeno Dynamics of microwave light. \textcolor{black}{Our} experiment shows that a mode that cannot \textcolor{black}{contain} $N$ photons \textcolor{black}{undergoes} a coherent evolution \textcolor{black}{that} is confined in phase space. Effectively, an electromagnetic mode is transformed into an $N/2$ spin. 
\textcolor{black}{As a result of this transformation,} exotic non-classical states develop under a simple resonant drive and share similar features with Schr\"odinger cat states of light. \textcolor{black}{Our} work could be extended to \textcolor{black}{allow} evolution outside of the exclusion circle, where the barrier acts as a peculiar scatterer in phase space~\cite{Raimond2012}. Other cat-like states and squeezed states could then be produced.
\textcolor{black}{Tailoring the Hilbert space in time is an important new resource for quantum control. Our experiment enables}
the realization of various protocols based on QZD, such as generation and protection of entanglement~\cite{Maniscalco2008,Wang2008,Shi2012}, and quantum logic operations~\cite{Shao2009}.  
\textcolor{black}{Moreover, by turning on and off in time the blockade level and} combining it with fast amplitude displacements, which is feasible by improving the cavity decay rate by an order of magnitude, it \textcolor{black}{would be} possible to realize phase space tweezers for light~\cite{Raimond2010,Raimond2012}. \textcolor{black}{A possible and important outcome would then be the manipulation of Schrödinger cat states in a unique way and the quantum error correction of cat-qubits, which are a promising quantum computing paradigm~\cite{Mirrahimi2014}.}


\begin{small}
~~\\
\noindent \textbf{\textup{\small Acknowledgments}}
We thank Michel Devoret, Sukhdeep Dhillon, \c{C}a$\mathrm{\breve{g}}$lar Girit, Takis Kontos, Zaki Leghtas, Vladimir Manucharyan, Mazyar Mirrahimi, Saverio Pascazio, the Quantronics Group, Jean-Michel Raimond, Pierre Rouchon, and J\'er\'emie Viennot. Nanofabrication has been made within the consortium Salle Blanche Paris Centre. This work was supported by the ANR contract ANR-12-JCJC-TIQS and the Qumotel grant Emergences of Ville de Paris. LB acknowledges support from Direction Générale de l'Armement.
\\

\noindent \textbf{\textup{\small Supplementary Materials}}
\\
www.sciencemag.org \\
Materials and Methods \\
Supplementary Text \\
Figs. S1 to S4 \\
References [26-32]

\end{small}

\end{document}